\documentclass[fleqn,12pt]{article}
\usepackage{amsmath,amssymb,epsfig,times,graphics}
\usepackage{amssymb}
\usepackage{amsmath}
\usepackage{CJK}
\usepackage{graphics}
\usepackage{graphicx}
\usepackage{dcolumn}
\usepackage{bm}
\usepackage{subfigure}
\usepackage{lineno}
\usepackage{amsthm}
\begin{document}

\title{\bf Control of epidemics via social partnership adjustment} \date{}
\maketitle

\author{Bin Wu$^{1}$, Shanjun Mao$^{2}$, Jiazeng Wang$^{3}$, Da Zhou$^{2,*}$}

\begin{enumerate}
  \item School of Sciences, Beijing University of Posts and Telecommunications, Beijing 100876, PR China
  \item School of Mathematical Sciences, Xiamen University, Xiamen 361005, PR China
  \\ ($*$Corresponding Author, zhouda@xmu.edu.cn))
  \item Department of Mathematics, Beijing Technology and Business University, Beijing 100048, PR China

\end{enumerate}

\section*{Abstract}
Epidemic control is of great importance for human society.
Adjusting interacting partners is an effective individualized control strategy.
Intuitively, it is done
either by shortening the interaction time between susceptible and infected individuals
or by increasing the opportunities for contact between susceptible individuals.
Here, we provide a comparative study on these two control strategies
by establishing an epidemic model with non-uniform stochastic interactions.
It seems that the two strategies should be similar,
since shortening the interaction time between susceptible and infected individuals
somehow increases the chances for contact between susceptible individuals.
However, analytical results indicate that the effectiveness of the former strategy sensitively depends on
the infectious intensity and the combinations of different interaction rates, whereas the
latter one is quite robust and efficient.
Simulations are shown in comparison with our
analytical predictions. Our work may shed light on the strategic choice of disease control.

\section{Introduction}
\label{}

Modeling the spreading of infectious diseases has a long history \cite{kermack1927contribution,anderson1992infectious,hethcote2000mathematics,
Wang:2014aa,faruque2005seasonal,da2006epidemics,boguna2013nature}.
Mathematical models not only deepen the understanding of epidemic dynamics, but also shed light on the control of diseases.
In recent years, much attention has been paid to the epidemic control via social relationship adjustment
\cite{gross2008adaptive,funk2010modelling,pastor2015epidemic}.
As a pioneering work, Gross \emph{et. al.} first proposed a susceptible-infected-susceptible (SIS) model on an adaptive network.
Therein the susceptibles break the link with the infected and rewire to another randomly selected susceptible individual \cite{gross2006epidemic}.
This rewiring rule brings in highly complex dynamics (such as bistability and oscillation) to  the classical SIS model.
The rewiring dynamics then opens the avenue on how individualized partnership adjustment alters the epidemic dynamics.
On the one hand,
besides the SIS model, typical epidemic models have almost been investigated
including susceptible-infective-recovered-susceptible (SIRS) model \cite{shaw2008fluctuating},
susceptible-infective-recovered (SIR) model \cite{lagorio2011quarantine} and susceptible-infective-vaccinated (SIV) model \cite{shaw2010enhanced}.
On the other hand, more realistic and complex link rewiring rules are proposed.
In particular,  generalizations of  Gross \emph{et al}'s rewiring rule are mainly  in two folds:
For one thing, after the disconnection
of susceptible-infected (SI) link, the susceptible is assumed to reconnect
to a randomly selected member of the population no matter it is susceptible or not \cite{zanette2008infection}.
For another, the infected is also allowed to switch its partnership from the susceptible to a new randomly selected contact \cite{risau2009contact}.
Besides the rewiring rule which is dependent on the infection process, the rewiring rule that is independent of the infection process
was also investigated \cite{fefferman2007disease,schwarzkopf2010epidemic}.
In spite of different model assumptions, all these models showed that,
the infection propagation can be greatly influenced by the dynamical networks. In particular, the infection
can be effectively suppressed by reducing the interaction opportunities between susceptible and infected individuals.

Besides the above-mentioned link-rewiring models, another type of adaptive networks is the link-activation-deactivation model
\cite{van2010adaptive,valdez2012intermittent,guo2013epidemic}.
It assumes that a link can either be broken or recreated on the basis of the infectious states
of the two endpoints of the link. In this model, only local information is required,
which could be more realistic \cite{pastor2015epidemic}.
In particular, Guo \emph{et. al.} proposed an ASIS model, in which any SI link can be broken (deactivated). After the disconnection
of an SI link, the two disconnected nodes can be reconnected again once both of them become susceptible (activated) \cite{guo2013epidemic}.
Despite of seemingly differences, the link-activation-deactivation dynamics is similar to the rewiring dynamics:
On the one hand, Guo \emph{et al} showed that the ASIS model (initiated on complete graphs) can approximate
the link-rewiring model in \cite{gross2006epidemic}.
On the other hand,
the quasi-stationary (metastable) fraction of infected individuals
can be reduced by increasing the effective breaking rate (proportional to the ratio of deactivating rate to activating rate).
This echoes the results based on the link-rewiring models that the disease can be controlled by reducing the contacts between susceptible and infected individuals.
Therefore, both types of the linking dynamics in epidemic control can be seen as decreasing the interaction rate between susceptible and infected individuals (called \emph{SI control}). Furthermore,
considering that the effective breaking rate in \cite{guo2013epidemic}
also depends on the activating rate between susceptibles,
their work reminds us of the significance of SS links in epidemic control.

Intuitively, increasing the interaction time between susceptibles can also be a control strategy (called \emph{SS control}).
Yet it is seldom addressed, compared to the SI control that has been intensively studied in previous literatures.
It seems that these two control strategies are the two sides of the same coin.
Actually, this is true in Gross's model \cite{gross2006epidemic},
since based on their rewiring rule, the decrease of SI links directly leads to the increase of SS links.
However, this is no longer valid in Risau-Gusm{\'a}n and Zanette's model \cite{risau2009contact},
since the disconnection of an SI link does not necessarily result in the reconnection
of an SS link.
Therefore the SS and SI control strategies are not equivalent in general.

In this work, we provide a comparative study on the SI control and the SS control by proposing a novel link-rewiring SIS model.
Unlike the models only allowing the breaking of SI links, we allow all the three types of links (SS, SI and II) to be broken, equipped
with three independent parameters to characterize the breaking rates of SS, SI and II links.
Actually, this assumption mimics the intrinsic nature of human mobility \cite{lydie2004mobility,mari2011modelling},
namely, people move or change their social relationships due to a variety of reasons, even without the consideration
of avoiding infectious diseases. In this way, SI links should not be the only type that is allowed to be broken,
both SS and II links can change. For example,
in \textbf{AIDS} (acquired immune deficiency syndrome) not only the susceptibles are willing to avoid contacts from
the infectives, but the susceptible-susceptible and infected-infected relationships
may also be broken up due to unsatisfactory sexual experiences, \emph{i.e.}
the rewiring processes can happen in SS and II links.
Besides, we allow all the individuals to be capable of adjusting any of their partners.
This mirrors the freedom of social life. It also excludes the central control of epidemics, for example, that via organizations.
In this way, we could concentrate on how the social partnership adjustment strategies alone alter the fate of epidemics.

We demonstrate analytically that our model captures the epidemic dynamics with non-uniform interaction rates under fast linking dynamics.
It is shown that sometimes the SS control is more effective and robust than the SI control.
In particular, strengthening the closeness between susceptibles (SS control) effectively eradicates the disease no matter how infectious the disease is.
However, the effectiveness of the SI control sensitively depends on the infectious intensity and the intrinsic mobility rate of the population.
In other words, there are cases such that the SI control cannot eliminate the disease so efficient as the SS control.
Simulation results are also shown for validating our theoretical predictions.
Our findings suggest that, besides the SI control, it is still of concern that the SS control may serve as a better candidate for epidemic control.

\section{Model and analysis}
In this section, we propose the model of epidemic spreading coupled with a simple stochastic link-rewiring dynamics.
Then we theoretically analyze the epidemic model with non-uniform interaction rates
 based on the time scale separation.

\subsection{Epidemic dynamics}
We consider a structured population of $N$ individuals.
The population is located on a connected network.
We assume that the average degree $k$ is much smaller than the population size $N$, \emph{i.e.} $k\ll N$.
Here nodes refer to individuals and links represent social ties between individuals.

We adopt a standard  susceptible-infected-susceptible (SIS) model to study the epidemic spreading.
The SIS model assumes that susceptible individuals get infected with a probability proportional to the number of their infected neighbors;
infected individuals recover and become susceptible with no immunity to the disease after a period of recovery time.
The SIS model has three features:
\emph{i}) the whole population size $N$ is constant over time;
\emph{ii}) the transmission of disease only happens via the SI links;
\emph{iii}) the recovery of infected individuals is independent of the status of their neighbors.

Let $I_t$ be the number of infected individuals at time $t$,
therefore, the mean-field equation of the SIS model on the structured population is given as follows
\begin{align}
\frac{d I_t}{dt}=\lambda N_{SI}-\mu I_t.
\label{BasicModel}
\end{align}
Here $\lambda$ is the transmission rate and $\mu$ is the recovery rate.
All through the paper we assume that $\mu=1$ without loss of generality,
and $N_{SI}$ is the number of the SI links.

\subsection{Link-rewiring dynamics}
The social relationships between individuals are not eternal,
but are continuously co-evolving.
As a typical example, susceptible individuals tend to avoid contacts with infected ones
by adjusting their local connections.
It has served as the most recognized prototype in the study of epidemic control
on dynamical networks. However, individuals may receive miscellaneous information
when making rewiring decisions, thus it is possible for all the individuals
to adjust all of their current social relationships.
Such a rewiring process captures mobility-like human behavior.

Here we propose a simple link-rewiring dynamics by extending the
dynamical nature from SI links to all types of links in the network.
Each individual is either susceptible (S) or infected (I).
Thus, there are three types of links: susceptible-susceptible (SS), susceptible-infected (SI) and infected-infected (II) links.
To characterize the fragilities of different types of links, we define $w_{XY}$
($XY\in\{SS, SI, II\}$) as the probability with which an $XY$ link breaks off in the process of disconnection.
In each rewiring step, a link $XY$ is selected randomly from the network.
With probability $w_{XY}$, the link is broken, otherwise the link remains connected.
If it is broken, $X$ or $Y$ is picked as the active individual, who is entitled to reform a new link. Its new neighbor is randomly selected from
the individuals who are not in its current neighborhood. Self-connections and double connections are thus not allowed here.

In this way, the link-rewiring dynamics can be modeled as a Markov chain in the state space of $\ \{SS, SI, II\}$ \cite{wu2010evolution,wu2011evolutionary,Wu20160282}.
Considering the transition probabilities between states,
let us take the transition from $SI$ to $SS$ as an example.
This happens only when $SI$ is broken off and $S$ is
selected to reform a new link to another susceptible individual. Note that the total population size
$N$ is much larger than the average degree $k$, the transition probability $Q_{SI\rightarrow SS}$ is approximately given by $w_{SI}\times \tfrac{1}{2} \times s$, where $s$ is the density of susceptibles at the moment.
Similarly, we calculate all the other
transition probabilities,
yielding the transition probability matrix
\begin{equation}
Q=\begin{array}{cccc}
 &\begin{array}{lll}
   SS & \ \ \ \  \ \ \ \ \ \ \  \ SI & \  \ \  \ \ \ \ \
\ \ \ \ \ \ II \end{array}\\
\begin{array}{c}SS\\SI\\II\end{array}&\left(\begin{array}{ccc}
$$sw_{SS}+(1-w_{SS})$$\ \ &$$iw_{SS}$$&\ \ $$0$$\\\ $$\tfrac{sw_{SI}}{2}$$\ \ &$$1-\tfrac{w_{SI}}{2}$$&\ \ $$\tfrac{iw_{SI}}{2}$$\\
$$0$$\ \ &$$sw_{II}$$&\ \ $$iw_{II}+(1-w_{II})$$
\end{array}\right)\end{array}\label{Matrix},
\end{equation}
where $i$ is the density of infected individuals. According to the standard theory of Markov chain \cite{gardiner1985handbook}, there exists a unique limiting distribution $\Pi$ satisfying $\Pi Q=\Pi$ provided $Q$ is irreducible and aperiodic. Namely, when $w_{SS}w_{SI}w_{II} i s\neq 0$, $Q$ has a unique stationary distribution
\begin{align}
\Pi &=(\pi_{SS}, \pi_{SI}, \pi_{II})\\ \nonumber
    &=\frac{1}{A(i)}\left(\frac{s^2}{w_{SS}}, \frac{2si}{w_{SI}}, \frac{i^2}{w_{II}}\right),
\label{stationary}
\end{align}
where $A(i)=\tfrac{s^2}{w_{SS}}+\frac{2si}{w_{SI}}+\frac{i^2}{w_{II}}$ is the normalization.

\subsection{Time scale separation}
It is challenging to capture $N_{SI}$ due to the complexity of real social
networks \cite{pastor2001epidemic,albert2002statistical,eubank2004modelling}.
This is already true in static networks, and it becomes even more difficult
taking into account the dynamical nature of social networks \cite{gross2008adaptive}.
Here we overcome this problem by
assuming the adiabatic elimination of fast linking dynamics \cite{gardiner1985handbook}
(also called annealed adaptive dynamics \cite{pastor2015epidemic}),
\emph{i.e.} the adjustment of social ties is much more frequent than the
update of infection states. This assumption implies \emph{time scale separation} of the two coupled dynamics.
In other words, the disease is unlikely to spread until the social configuration tends to the stationary regime.
In this way, $N_{SI}$ is approximated as
\begin{equation}
N_{SI}\approx\frac{Nk}{2}\pi_{SI},
\label{Nsi}
\end{equation}
where $Nk/2$ is the total number of the links in the network and $\pi_{SI}$ is the fraction of SI links in the stationary regime. This approximation greatly reduces the complexity of the coupled dynamics.
In light of this, the idea of time scale separation has frequently been used in analyzing complex dynamics on adaptive networks
(epidemics \cite{schwarzkopf2010epidemic,van2010adaptive,guerra2010annealed},
evolutionary games \cite{wu2010evolution,pacheco2006coevolution}).

By taking Eq. (\ref{Nsi}) into Eq. (\ref{BasicModel}) we have
\begin{equation}
\frac{d I}{dt}=\frac{\lambda Nk}{2}\pi_{SI}-\mu I.
\label{eq:model22}
\end{equation}
Note that $i=I/N$, $s=S/N$, and $\pi_{SI}=\tfrac{2is}{A(i)w_{SI}}$, Eq. \eqref{eq:model22} can be transformed to
\begin{equation}
\frac{d i}{dt}= \frac{k\lambda}{A(i) w_{SI}}i s-\mu i.
\label{model}
\end{equation}
In particular, when all the interaction rates are uniform and positive
($w_{SS}=w_{SI}=w_{II}>0$), Eq. (\ref{model}) reduces to
\begin{equation}
\frac{d i}{dt}= i\left(\underbrace{k\lambda}_{\lambda_e} s-\mu \right).
\label{BasicModel2}
\end{equation}
Eq. (\ref{BasicModel2}) is nothing but the classical SIS model \cite{anderson1992infectious},
provided that $\lambda_e= k\lambda$ is redefined as the effective transmission rate.
This implies that the population is as if a well-mixed population,
if individuals break their partnerships with no social bias.
It should be pointed out that, when $w_{SS}=w_{SI}=w_{II}=0$,
the transition probability matrix Eq. (\ref{Matrix}) violates the irreducible condition \cite{durrett2010probability} that our analysis replies on.
In fact, this case resembles the static network, which has been excluded from our analysis.

When the interactions are violated from above social unbias,
on the one hand,
it results in non-uniform interactions in the population \cite{taylor:TPB:2006}.
Therefore, Eq. (\ref{model}) extends the classical SIS model
from uniform interaction rates to non-uniform interaction rates.
Noteworthy, this non-uniform extension is an emergent property from microscopic stochastic linking dynamics,
which is not assumed in prior.
On the other hand, if we define $\Lambda(i)=\frac{k\lambda i}{A(i) w_{SI}}$, our model also extends the classical SIS model
from density independent transmission rate to density dependent transmission rate \cite{gao1992disease,zanette2008infection}.
In other words, the dynamical nature of social networks essentially acts as a feedback mechanism on the SIS model.
The feedback mechanism, which is taken as the central idea of control, can significantly alter the epidemic dynamics.

Noteworthy, all the analysis above are based on the time scale separation.
Thus it suggests that the link-rewiring event should happen with a sufficiently large probability
(close to 1) in each update.
Furthermore, we give a more precise lower bound for this probability based on pair approximations:
It is found that the time scale separation is at work
provided the likelihood of the linking dynamics is greater than (see Appendix \ref{A})
\begin{align}
\frac{k^2}{k^2+N}.
\end{align}
For more general cases where the time scale separation is absent,
higher order approximation method could be applied to provide theoretical
insights \cite{vazquez2015rescue} (see Appendix \ref{A}).

\section{Theoretical Results}
\label{results}

Our main concern in this comparative study is
epidemic control via changing the interaction rates in different ways.
Based on Eq. (\ref{BasicModel}), it is $N_{SI}$ that determines the spread of infection.
The more the SI links are, the more likely the spread of infection could be.
Generally, there are two ways to control $N_{SI}$. For one thing,
it is natural to increase $w_{SI}$ for reducing the interaction rate ($1/w_{SI}$) between susceptible and infected individuals (\emph{SI control}).
For another thing, decreasing $w_{SS}$ can also reduce the exposure of susceptibles to infection (\emph{SS control}).
Therefore, we will investigate the control of epidemics via these two strategies.
More specifically, by taking the uniform interaction rates ($w_{SS}=w_{SI}=w_{II}>0$) as the reference case,
we would like to provide a comparative study on both the SI control ($w_{SI}>w_{SS}=w_{II}$) and the SS control ($w_{SS}<w_{SI}=w_{II}$).
In the following, we assume that the effective transmission rate is always larger than the recovery rate,
\emph{i.e.} $\lambda_e>1$, where the epidemic control is necessary.

\subsection{SI control: Decreasing the interaction rate between susceptible and infected individuals}
\label{SI}

\begin{figure}[htbp]
   \centering
   \includegraphics[width=6in]{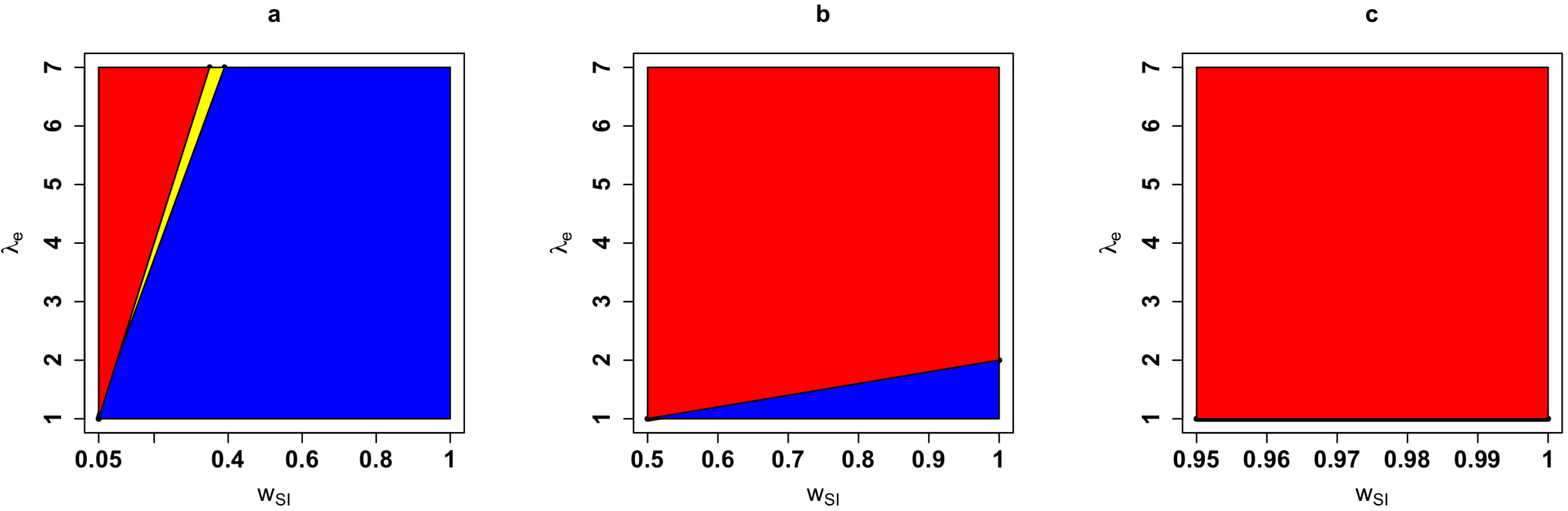}
\caption{SI control of epidemics, \emph{i.e.}, increasing the breaking probability between susceptible and infected individuals.
For the uniform interaction case ($w_{SS}=w_{SI}=w_{II}$),
the model degenerates to the conventional SIS model.
There is only one internal equilibrium and it is stable provided $\lambda_e>1$.
Here we solely adjust $w_{SI}$ such that the duration time of $SI$ links is shorter than the other two types of links,
\emph{i.e.} $w_{SI}>w_{II}=w_{SS}$. The three panels show the phase diagrams in the $(w_{SI}, \lambda_e)$-plane.
The quality of the SI control is sensitively dependent on the reference uniform breaking probabilities.
(a) When they are small ($w_{SS}=w_{II}=0.05$), decreasing the interaction between susceptible and infected individuals makes the
phase diagram change from endemic state (red) to bistable state (yellow) and then to final extinct state (blue).
(b) When $w_{SS}=w_{II}=0.5$, there is no bistable state (yellow) any more.
This implies it becomes hard to eradicate disease when the population is even more mobile.
(c) The right panel shows that the SI control is incapable of eradicating the disease provided
the population is intrinsically highly mobile ($w_{SS}=w_{II}=0.95$).
}
 \label{fig:wss=wii}
\end{figure}

To decrease the interaction rate between susceptibles and infectives, it is equivalent to increase the breaking probability $w_{SI}$.
Based on the uniform interaction as the reference case, we are interested in how the epidemic dynamics is changed by increasing $w_{SI}$.
Here the uniform interaction can mimic the basic migration rate in the population.
To illustrate our main results, we consider three typical cases with different initial values of the
uniform interaction rates (see Appendix \ref{Tech} for technical details):

\emph{Small initial case (Fig. \ref{fig:wss=wii}a)}. In this case, we set initially the breaking probabilities for all types of links to be $0.05$. The disease can be controlled by increasing $w_{SI}$ from 0.05 to 1. In particular, for small infectious rate (\emph{i.e.} $\lambda_e\leq2$), there is a phase transition with the increase of $w_{SI}$. That is, the final state of epidemics turns from endemic to extinction. For large $\lambda_e$ (\emph{i.e.} $\lambda_e>2$), there is a small region of bistability where the disease persists or die out due to the initial infected fraction. Compared to the single continuous phase transition in the conventional (uniform) SIS model, the non-uniform SIS model can give rise to multiple phase transitions.
The emergent bistability in adaptive SIS model has already been reported in previous studies
\cite{gross2006epidemic,zanette2008infection,graser2011separatrices,tunc2014effects}, but it is quite difficult to approximate the conditions under which bistability is present.
For our model, we explicitly provide those analytical conditions under which the bistability emerges based on Eq. \eqref{model}.
In the case of SI control
($w_{SI}>w_{SS}=w_{II}$), it arises if and only if
\begin{align}
(\lambda_e, w)\in(2, +\infty)\times\left(\lambda_e, \frac{\lambda_e+\sqrt{(\lambda^2_e+(\lambda_e-2)^2)}}{2}\right),
\label{eq:0000}
\end{align}
(see Appendix \ref{Tech}) where $w=w_{SI}/w_{SS}$.

\begin{figure}[htbp]
   \centering
   \includegraphics[width=4in]{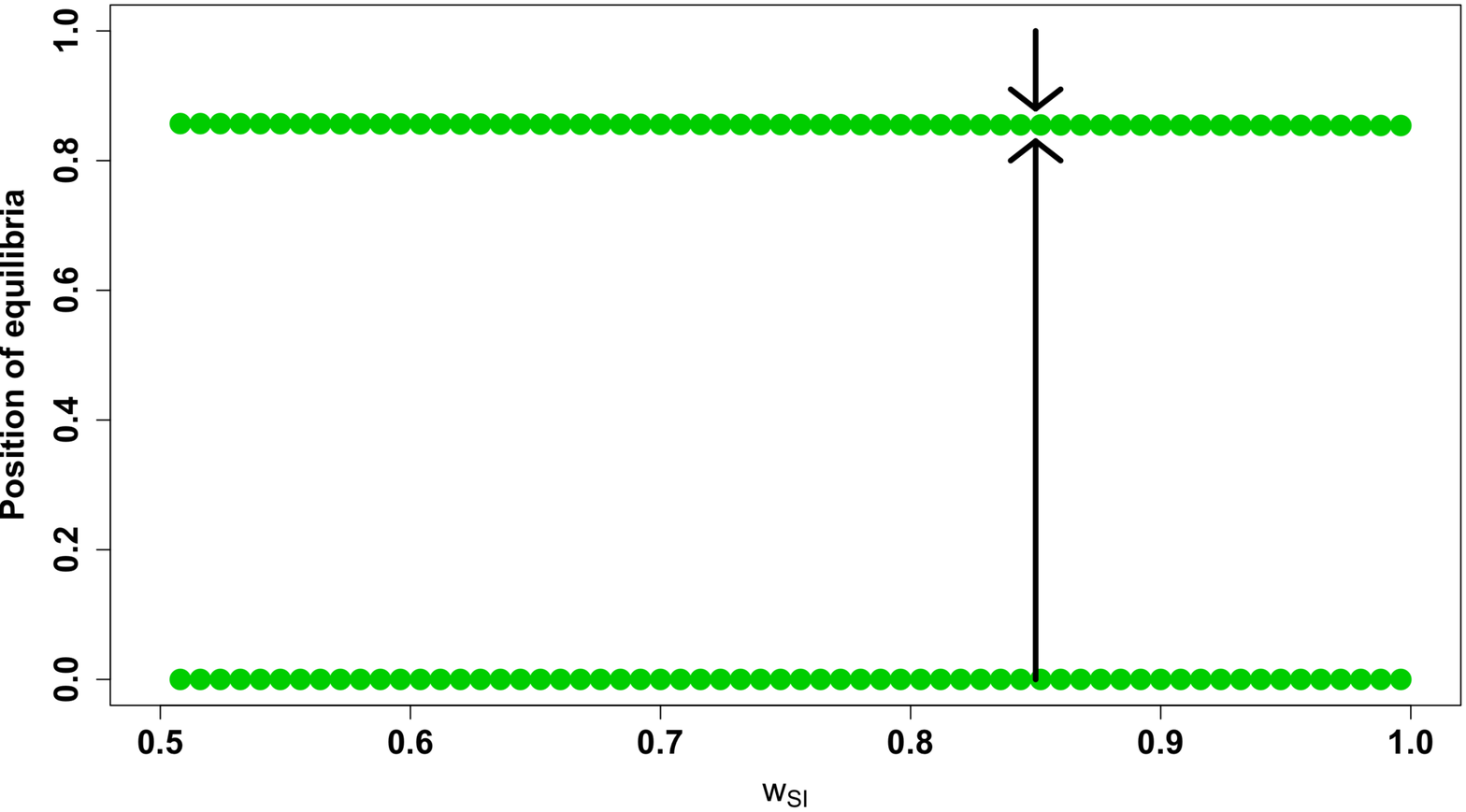}
\caption{Equilibrium position of infected fraction as a function of $w_{SI}$.
Here $w_{SS}=w_{II}=0.5$, and $\lambda_e=7$. The disease cannot be eradicated by the SI control,
and the level of infection in the equilibrium state declines very slowly (from $0.857$ to $0.854$) by
increasing $w_{SI}$ (from $0.5$ to $1$).
}
   \label{fig:2}
\end{figure}

\emph{Intermediate initial case (Fig. \ref{fig:wss=wii}b)}. In this case, increasing $w_{SI}$ is not as effective as that in the above small initial case.
For small $\lambda_e$, even though there still exists a phase transition from endemic state to extinct state, the marginal value of $w_{SI}$ that needs to cross the transition line is large.
More importantly, when $\lambda_e$ is large enough, increasing $w_{SI}$ is unable to eradicate the disease any more. The disease
will persist no matter how large the interaction rates between susceptibles and infectives are.
Moreover, it is shown in Fig. \ref{fig:2} that, the endemic level is not sensitive to $w_{SI}$. In other words,
by increasing $w_{SI}$, the final fraction of infectives declines very slowly.
That is to say, the increase of $w_{SI}$ can neither qualitatively change the final state of endemic, nor
quantitatively inhibits the final fraction of infectives.

\emph{Large initial case (Fig. \ref{fig:wss=wii}c)}. In this case, the endemic state is always the global stable state
provided $\lambda_e>1$. That is, the epidemics cannot be eradicated by the SI control.

To summarize, the control efficiency via reducing the interaction rate between susceptibles and infectives strongly depends on the reference breaking probabilities, \emph{i.e.}, the intrinsic population mobility.
The more likely the population is mobile, the worse the SI control performs.

\subsection{SS control: Increasing the interaction rate between susceptibles}
\label{SS}

Unlike the SI control, increasing the interaction rate between susceptibles is shown as an effective and robust strategy
for epidemic control. In fact, no matter what the intrinsic mobility rate of the population is,
the SS control successfully eradicates the disease.
To this end, we study the three typical reference population mobility cases in the above subsection (see Appendix \ref{Tech} for technical details).
Fig. \ref{fig:wsi=wii} shows that the phase diagrams for the three cases are quite similar to each other:

\begin{itemize}
  \item For small $\lambda_e$ ($1<\lambda_e\leq2$), by decreasing $w_{SS}$, the final state of disease is directly transformed from endemic to extinction.
  \item For large $\lambda_e$ ($\lambda_e>2$), the bistablilty arises in all the three cases. That is, no matter how large the initial uniform interaction rates are,
  with the decrease of $w_{SS}$, there is an intermediate region where the disease persists or dies out depending on the initial fraction of disease
  Furthermore, we analytically obtain that the bistable region is given by
  \begin{align}
  \left(\lambda_e, \frac{1}{w}\right)\in(2, +\infty)\times\left(\frac{4}{4+\lambda^2_e}, \frac{1}{\lambda_e}\right).
  \label{eq:000}
  \end{align}
\end{itemize}

By comparison, the SS control is more effective than the SI control in two ways. On one hand, the control of $w_{SS}$ is independent of the
intrinsic population mobility, \emph{i.e.}, robust control. On the other hand, decreasing $w_{SS}$ can always effectively eradicate the disease regardless of infectious intensity
(Fig. 4 illustrates the position of equilibria as a function of $w_{SS}$ in the bistable case).

\section{Agent-based simulations}

In this section, we present agent-based simulations and further discuss the efficiency of the time scale separation method
based on the comparison between the simulation results and theoretical predictions.

\subsection{Simulation procedures}
The \emph{contact process} \cite{liggett2013stochastic} is adopted to model the epidemic spreading
on networks.
Let $\alpha\in(0,1)$ be the probability of epidemic spreading in each update.
The simulation is performed as following:

\begin{enumerate}
\item Initially, there are $N$ individuals located on a regular graph with degree $k$,
where each individual has exactly $k$ neighbors.
Then $N_0$ infectives and $N-N_0$ susceptibles are randomly distributed.
\item Generate a random number $r\in(0,1)$. If $r<\alpha$, we perform the contact process. Otherwise ($r\geq\alpha$), we perform the linking dynamics.
\item If the contact process occurs, an infected individual (called Bob) is selected randomly.
With probability $\frac{\mu}{k_{\text{Bob}}\lambda+\mu}$ Bob becomes susceptible,
where $k_{\text{Bob}}$ is the degree of Bob.
Otherwise a neighbor of Bob's is selected at random.
This neighbor, namely Jack, is infected with probability $\frac{\lambda}{k_{\text{Bob}}\lambda+\mu}$.
Noteworthy, Jack becomes infected if its status is susceptible.
However, this new infection event does not change the state of Jack if Jack has been infected already.
Then return to Step 2.
  \item If the linking dynamics occurs, a link is selected randomly.
  The type of this link is denoted as  $XY$ ($XY\in \{SS, SI, II\}$).
    With probability $w_{XY}$, the link is broken,
  otherwise the link remains connected. If it is broken, $X$ or $Y$ is picked as the active individual, who is entitled to reform a new link.
  The new neighbor is randomly selected from the individuals who are not in its current neighborhood. Then return to Step 2.
\end{enumerate}
Each data point is averaged over $50$ independent samples.
In each sample, we run a transient time of $10^6$ generations,
and we set the mean value over time window of last $10^3$ generations to be the final fraction of infectives.

It should be pointed out that, the simulation results are robust for all initial connected graphs,
provided the number of infectives $N_0$, population size $N$ and the average degree $k$ are fixed.
The regular graph here only serves as a prototype for simulations.
In fact, our linking dynamics is a Markov chain, which is irreducible and aperiodic.
This yields that the limiting behavior is independent of the initial configuration of the network \cite{durrett2010probability}.
Furthermore, the assumption of time scale separation allows all the links to converge to the stationary distribution. Therefore, all the links would converge to the stationary distribution no matter what type of graph it is initially.

\subsection{Simulation results}
With the coupled linking dynamics,
the final fate of the infection can be of three folds:
die out no matter what the initial fraction of the infective is (called \emph{extinction});
stabilize at a non-zero fraction of infectives no matter what the (positive) initial fraction of infectious individuals is (called \emph{endemic});
stabilize at a non-zero fraction of infectives if the initial fraction of infectious individuals exceeds a critical value
and die out otherwise (called \emph{bistability}).

For the extinction cases,
simulation results are found to be in good agreement with the analytical predictions.
This is true for all the parameter regions predicting extinction for both SI and SS controls (see Fig. \ref{fig:si:1}).

For the endemic cases,
Fig. \ref{fig:si:3} shows that
the population would end up with a constant fraction of infected individuals,
provided there are infective individuals initially.
This is exactly in line with the analytical predictions.
Furthermore, the inconsistency between the analytical and simulation results is less than $10\%$,
which is acceptable. Considering this $10\%$ disagreement,
the analytical predictions systematically over-estimate the simulation results.
In fact, the agent-based contact process is a Markov process with an absorbing state,
where no infected individual is present.
In other words, the disease would go extinct eventually if the system evolves sufficiently long.
Our analytical results, however, are in the quasi-stationary time scale \cite{naasell1999quasi,zhou2010evolutionary}.
The inconsistency between the analytical and simulation results suggests that
the  running time $10^6$ is beyond the quasi-stationary time scale.
Thus the system may evolve to the absorbing state with non-negligible chances.

For the bistability cases,
the simulation results show qualitative agreement with the analytical predictions.
In particular,
the critical initial fraction of infected individuals, ensuring a dramatic outbreak of epidemics, is consistent with the
unstable fixed point predicted by the analytical result (see the blue dash lines in Fig. \ref{fig:si:2}).
Disagreements, however, are also present.
For example, the theoretical results tend to underestimate the final infection when the infection fraction is rare initially.
In fact, this bistable case bears two internal equilibria lying at $x_1^*$ (unstable) and $x_2^*$ (stable) ($x_1^*<x_2^*$).
For small initial fraction of infectives,
the deterministic part of the system drives the infection to extinction based on the analytical investigation.
Yet by its intrinsic stochastic nature of the epidemic spreading,
the infection would increase in number and be possibly trapped around the stable equilibria from time to time.
Even though it is a type of \emph{rare event}, it takes quite long to escape from this trap.
Thus on average it results in a relatively higher level of final fraction of infectives given the running time of
simulations (here $10^6$ generations).
In other words, it is the  interplay between the stochastic effect and stable
equilibrium at zero that results in such inconsistency.
Noteworthy, despite of this quantitative inconsistency,
the salient feature of the bistable dynamics is still captured by the analytical predictions.

In Fig \ref{fig:si:4}, we investigate how the population size
affects the accuracy of the analytical approximation.
Theoretically,
large population size inhibits the stochasticity arising
from the finite population effect,
which is closer to the mean-field approximation.
Similar discussions can be found in \cite{wu2010evolution}.
Fig. \ref{fig:si:4} shows
the case with $N=100$ still captures the bistable dynamics as the case with $N=500$ does.

\section{Discussions and Conclusions}

We have proposed a simple link-rewiring rule to model social partnership adjustment.
Therein all the links are about to break,
capturing the mobility nature of the population.
This simple model paves the way to compare different rewiring-based epidemic control strategies.

Instead of focusing on the control strategy via breaking SI links (\emph{e.g.} \cite{gross2006epidemic}),
our model extends the rewiring rule from SI links to all the three types (SI, SS and II) of links,
which facilitates us to compare different rewiring control strategies.
We find that, for mild infectious disease, both SI and SS control strategies can eradicate the disease.
For strong infectious disease, however, it is more efficient to adopt the SS control than the SI control.
This result is counterintuitive.
Intuitively, reducing the contacts between susceptible and infected
individuals is believed to suppress the disease propagation. Moreover, it seems that decreasing the interaction
rate of SI links could naturally result in the increase of SS links \cite{graser2011separatrices}.
How can these two strategies perform so differently?
One of the salient features of our model is the variability
of II links, which is seldom addressed previously.
 Actually,
increasing $w_{SI}$ is equivalent to decreasing both $w_{SS}$ and $w_{II}$.
In other words, the SI control is equivalent to simultaneously strengthening
SS links and II links. Similarly, the SS control is equivalent to simultaneously
reducing the closeness of SI links and II links. Thus, the relation of the SI and SS
control strategies is not as straightforward as expected.
To illustrate the impact
of II links on the epidemic dynamics, we consider two examples: (1) $w_{SI}=0.98$, $w_{SS}=0.2$,
$w_{II}=0.2$; (2) $w_{SI}=0.98$, $w_{SS}=0.2$, $w_{II}=0.98$.
The only difference between these two examples is the value of $w_{II}$.
It is easy to show that in example (1) disease becomes extinct,
whereas bistability arises in example (2) (based on Eq. \eqref{eq:000}).

Another feature of our reconnection rule is \emph{nonselective}.
In other words,
individuals are allowed to rewire to a randomly selected member no matter it is susceptible or not.
Compared to the \emph{selective} rule in \cite{gross2006epidemic} (rewiring to a randomly selected susceptible),
individuals in our model are not necessary to know who gets infected currently, which is more realistic.
Actually, the nonselective rule increases the exposure of the susceptible to the infected.
This is very likely in the beginning of epidemic season,
where the information on infection status is unaccessible.
In particular, even though the SI control increases the breaking possibility of each SI link,
a new SI link may be generated again due to the nonselective rule.
By contrast, the SS control
makes a straightforward intervention during the process of disconnection.
That is,
by strengthening the closeness between susceptibles, the SS strategy reduces the possibility of SI connection
effectively. In this way, the nonselective rule has a relatively small impact on the SS control. Therefore,
in the framework of the nonselective rewiring rule, the SS strategy is more efficient than the SI strategy.

Concentrating on the relation between the lifespan
of each type of links and epidemic spreading, our model does not account for
other features that are also considerable in capturing the epidemic dynamics of real world
networks. For example, (1) our linking dynamics does not take into account the social interactions with memory,
such as friendship and working partners, in which individuals preserve the contacts that they used to make
\cite{valdez2012intermittent,guo2013epidemic};
(2) The link-rewiring process is a strong simplification of the real adaptive networked human behavior.
It is not necessarily realistic for individuals who break up a relationship to have a new partner immediately.
However, it probably mimics the dynamics of networks in \textbf{AIDS} to some extent:
The susceptible individuals break up their (mostly sexual) relationships with their infected partners and switch to other
perceived healthy individuals. Moreover, the infected individuals may also rewire their links to
other infectives.

To sum up,
our result captures the causation between the link fragility and the disease control.
Furthermore, this model might serve as a starting point to compare different
rewiring control strategies for more general models closer to reality.


\appendix

\section{Another analytical approximation}
\label{A}
Our model couples the linking dynamics and the epidemic dynamics.
While the method in the main text is analytically insightful,
it requires the time scales of the two dynamics to be separated.
In other words,
individuals should adjust their partners much faster than the spread of epidemics  to make this method applicable.
This is, however, not the case in general.
We propose another analytical method to overcome this restriction.
The method is based on pair approximation and rate equations \cite{vazquez2015rescue}.
Here we concentrate on how the method helps us estimate the condition
under which the time scale separation is valid.

Let $<I>$ and $<S>$ be the global frequencies of infected and susceptible individuals \emph{i.e.} $i$ and $s$ in Eq. \eqref{model};
and let $<XY>=N_{XY}/N$ be the frequencies of $XY$ pairs,
where $XY\in\{II,SI,SS\}$.
Thus $ <I>+ <S> = 1$ and  $<II>+<SI>+<SS>= 1$ hold.
The system thus is determined by three independent variables: $<I>$, $<SI>$ and $<II>$.
The crucial assumption for pair approximation is that higher-order of moments can be captured by moments of pairs.
In the following, we write down the rate equations of the three variables under the assumption of pair approximation.

For the evolution of fraction of the infected,
it is only determined by the epidemic dynamics.
In this case, the number of infected individuals increases or decreases by one, or stays the same in one time step.
By the Kolmogorov Forward Equation, we have that
\begin{align}
\Delta <I>=\text{Prob}\left(\Delta <I>=\frac{1}{N}\right)\frac{1}{N}
-\text{Prob}\left(\Delta <I>=-\frac{1}{N}\right)\frac{1}{N}.
\label{eq:001}
\end{align}
In particular, the probability that infected individuals increase by one in number happens:
1) the epidemic spreading is ongoing (with probability $\alpha$);
2) a susceptible individual is selected (with probability $<S>$), and it is infected by one of its infected neighbors.
The fraction of the infected individuals around a susceptible individual is $\tfrac{<SI>}{<S>}$ based on pair approximation.
Thus there are on average $\tfrac{k<SI>}{<S>}$ infected neighbors around the selected susceptible individual,
where $k$ is the average degree of the entire network.
Therefore, the infection probability of the susceptible within a small time interval $\Delta t$ is $\lambda \tfrac{k<SI>}{<S>}\Delta t$.
Thus $\text{Prob}\left(\Delta <I>=\frac{1}{N}\right)=\alpha\lambda k<SI>\Delta t$.
Similarly, we have  $\text{Prob}\left(\Delta <I>=-\frac{1}{N}\right)=\alpha\mu<I>\Delta t$.

Let us rescale the time interval  $\Delta t=1/N$.
For large population size $N$,
dividing $\Delta t$ on both sides of Eq. \eqref{eq:001} yields
\begin{align}
\dot{<I>}&=\frac{\alpha}{N}\left(\lambda k <SI>-\mu <I>\right).
\label{eq:02}
\end{align}
This equation is identical with the mean-field SIS model Eq. \eqref{BasicModel}  up to a rescaling factor.

For the evolution of the links, it can be caused both by the linking dynamics and the epidemic spreading.
Taking the change of $<II>$ as an example:
When the linking dynamics happens (with probability $1-\alpha$),
$II$ links would increase by one if an $SI$ link is selected, then broken, and the infected individual of the $SI$ link is selected,
and it switches to another infected individual (with probability $\frac{<I><SI>w_{SI}}{2}$),
$II$ links would decrease by one if an $II$ link is selected, then broken, and the selected infected individual switches to a susceptible individual  (with probability $<S><II>w_{II}$);
When the epidemic spreading happens (with probability $\alpha$):
For the recovery event,
an infected individual  is selected (with probability $<I>$),
it recovers with probability $\mu\Delta t$.
If the selected $I$ individual has $q$ ($0\leq q\leq k$) infected neighbors (with probability ${k \choose q} (\tfrac{<II>}{<I>})^q (1-\tfrac{<II>}{<I>})^{k-q}$),
the change of $II$ links is $-q$;
For the infection event,
a susceptible individual is selected (with probability $<S>$),
if it has $h$ ($0\leq h\leq k$) infected neighbors (with probability ${k \choose h} (\tfrac{<SI>}{<S>})^h (1-\tfrac{<SI>}{<S>})^{k-h}$),
the infection happens with probability $h \lambda \Delta t$,
and the change of $II$ links in this case is $h$.
Taking into account the formula of the expectation and the variance of the binomial distribution yields
\begin{align}
\label{eq:LinkEvo-011}
\dot{<II>}
&=(1-\alpha)\left\{\frac{<I><SI>w_{SI}}{2}-(1-<I>)<II>w_{II}\right\}\frac{2}{k}\nonumber\\
&+\frac{2\alpha}{N}\left\{\mu<II> +\lambda <SI> +\lambda (k-1)\frac{<SI>^2}{1-<I>}\right\}.
\end{align}

With similar arguments we have
\begin{align}
\dot{<IS>}&=\frac{2(1-\alpha)}{k}\left\{<II>w_{II}\left(1-<I>\right)+(1-<II>-<IS>)w_{SS}<I>-\frac{1}{2}w_{IS}<IS>\right\}\nonumber\\
&+\frac{2\alpha\mu}{N}\left(2<II>-\frac{<I>}{k}\right)
+\frac{2\alpha(k-1)\lambda}{N}<SI>\left(1+\frac{<SI>}{1-<I>}\right).
\label{eq:LinkEvo-03}
\end{align}

Finally we obtain the equations of moments with closed forms, \emph{i.e.}, Eqs. \eqref{eq:02}, \eqref{eq:LinkEvo-011} and \eqref{eq:LinkEvo-03}.
This method has been used in both evolutionary game theory \cite{ohtsuki2006simple} and
epidemic dynamics \cite{pastor2015epidemic,vazquez2015rescue} before.
These equations can be employed to investigate the coupled dynamics of links and epidemics for any time scales.


Furthermore,
the dynamics of $<SI>$ and $<II>$ can help us figure out the condition under which the time scale separation is valid.
The time scale separation requires that the evolution of links is mainly determined by the link-rewiring process.
It implies that $\tfrac{2\alpha(k-1)\lambda}{N}\ll \frac{2(1-\alpha)}{k}$ and $\tfrac{2\alpha\mu}{N}\ll \frac{2(1-\alpha)}{k}$ based on Eqs. \eqref{eq:LinkEvo-011} and \eqref{eq:LinkEvo-03}.
Let us assume that both the infection rate $\lambda$ and the recovery rate $\mu$ are of order one.
Then the two inequalities implies
\begin{align}
\alpha\ll \left(\frac{k^2}{N}+1\right)^{-1}.
\label{eq:criterion}
\end{align}
This necessary condition is a more precise criterion compared with $\alpha\ll 1$ to ensure the time scale separation.
It suggests that the condition for the time scale separation would be
more demanding with the increasing of the average degree $k$.
This also supports our assumption in the main text that $k$ should be much smaller than $N$.

\section{Dynamical analysis}
\label{Tech}

Here we give a rigorous dynamical analysis on Eq. (\ref{model}),
based on which the main results in Sec. \ref{results} are obtained.
Rewriting Eq. (\ref{model}) leads to
\begin{align}
\frac{di}{dt}= \frac{1}{A(i)} f(i),
\end{align}
where the cubic polynomial $f(i)$ is given by
\begin{equation}
(2w_{SS}w_{II}-w_{SI}w_{II}-w_{SS}w_{SI})i^3+(2w_{SI}w_{II}-(2+\lambda_e)w_{SS}w_{II})i^2
+(\lambda_e w_{SS}w_{II}-w_{SI}w_{II})i.
\label{gradient}
\end{equation}

The asymptotic properties of Eq. (\ref{model}) are totally determined by $f(i)$, since $A(i)$ is positive.
Note that $f(0)=0$, $i=0$ is a fixed point.

When $2w_{SS}w_{II}-w_{SI}w_{II}-w_{SS}w_{SI}=0$,
$$f(i)=iw_{II}[(2w_{SI}-(2+\lambda_e)w_{SS})i+(\lambda_e w_{SS}-w_{SI})].$$
\begin{itemize}
  \item If $\lambda_e w_{SS}-w_{SI}\leq0$, $i=0$ is the only stable fixed point. The infection will finally die out;
  \item If $\lambda_e w_{SS}-w_{SI}>0$, $i=0$ is an unstable fixed point,
  and
$$i=\frac{w_{SI}-\lambda_e w_{SS}}{2w_{SI}-(2+\lambda_e)w_{SS}}$$
becomes the only stable fixed point, corresponding to endemic infection.
\end{itemize}
It is shown that there exists a phase transition at $\lambda_e=w_{SI}/w_{SS}$, which is quite similar to the conventional SIS model
in which the critical point is located at $\lambda=1$.

When $2w_{SS}w_{II}-w_{SI}w_{II}-w_{SS}w_{SI} \neq 0$, it is possible for the model to give rise to bistability.
Let $w=w_{SI}/w_{SS}$, we have
\begin{itemize}
  \item If $w_{SS}=w_{II}$, bistable $\Leftrightarrow$ $\left(\lambda_e, w\right)\in(2, +\infty)\times\left(\lambda_e, \frac{\lambda_e+\sqrt{\lambda^2_e+(\lambda_e-2)^2}}{2}\right)$;
  \item If $w_{SI}=w_{II}$, bistable $\Leftrightarrow$ $(\lambda_e, \tfrac{1}{w})\in(2, +\infty)\times\left(\frac{4}{4+\lambda^2_e}, \frac{1}{\lambda_e}\right)$.
\end{itemize}
To show how we get the above results, we take the case $w_{SS}=w_{II}$ as an example. In this case,
$$f(i)=2w^2_{SS}i\underbrace{[(1-w)i^2+(2w-(2+\lambda_e))i+(\lambda_e-w)]}_{g(i)},$$
and its discriminant is denoted as
$\Delta=(2w-(2+\lambda_e))^2-4(1-w)(\lambda_e-w)$,
then the sufficient and necessary condition for bistability is given by
\begin{equation}
  \begin{cases}
    g(0)<0,\\
    g(1)<0,\\
    \Delta>0,\\
    0<-\frac{2w-(2+\lambda_e)}{2(1-w)}<1.
  \end{cases}
\end{equation}
By solving the above set of inequalities, we obtain that $\left(\lambda_e, w\right)\in(2, +\infty)\times\left(\lambda_e, \frac{\lambda_e+\sqrt{\lambda^2_e+(\lambda_e-2)^2}}{2}\right)$.
Similarly, we get the result for the case $w_{SI}=w_{II}$.

\section*{Acknowledgements}

We thank the referees for their helpful comments. Discussions with Prof. Ming Tang are greatly acknowledged.
D.Z. is grateful for funding by the National Natural Science Foundation of China (No. 11401499),
the Natural Science Foundation of Fujian Province of China (No. 2015J05016), and the Fundamental
Research Funds for the Central Universities in China (Nos. 20720140524, 20720150098).
B.W. is grateful for funding by the National Natural Science Foundation of China (No. 61603049), and
the Fundamental Research Funds for the Central Universities.


\begin{thebibliography}{10}

\bibitem{kermack1927contribution}
William~O Kermack and Anderson~G McKendrick.
\newblock A contribution to the mathematical theory of epidemics.
\newblock In {\em Proc. Roy. Soc. London Ser. A}, volume 115, pages 700--721.
  The Royal Society, 1927.

\bibitem{anderson1992infectious}
Roy~M Anderson and Robert~M May.
\newblock {\em Infectious Diseases of Humans: Dynamics and Control}.
\newblock Oxford University Press, Oxford, 1991.

\bibitem{hethcote2000mathematics}
Herbert~W Hethcote.
\newblock The mathematics of infectious diseases.
\newblock {\em SIAM Rev.}, 42(4):599--653, 2000.

\bibitem{Wang:2014aa}
Wei Wang, Ming Tang, Hui Yang, Younghae Do, Ying-Cheng Lai, and GyuWon Lee.
\newblock Asymmetrically interacting spreading dynamics on complex layered
  networks.
\newblock {\em Sci. Rep.}, 4:5097, 2014.

\bibitem{faruque2005seasonal}
Shah~M Faruque, Iftekhar~Bin Naser, M~Johirul Islam, ASG Faruque, AN~Ghosh,
  G~Balakrish Nair, David~A Sack, and John~J Mekalanos.
\newblock Seasonal epidemics of cholera inversely correlate with the prevalence
  of environmental cholera phages.
\newblock {\em Proc. Natl. Acad. Sci. USA}, 102(5):1702--1707, 2005.

\bibitem{da2006epidemics}
MM~Telo Da~Gama and A~Nunes.
\newblock Epidemics in small world networks.
\newblock {\em Eur. Phys. J. B: Condensed Matter}, 50(1-2):205--208, 2006.

\bibitem{boguna2013nature}
Marian Bogu{\~n}{\'a}, Claudio Castellano, and Romualdo Pastor-Satorras.
\newblock Nature of the epidemic threshold for the
  susceptible-infected-susceptible dynamics in networks.
\newblock {\em Phys. Rev. Lett.}, 111(6):068701, 2013.

\bibitem{gross2008adaptive}
Thilo Gross and Bernd Blasius.
\newblock Adaptive coevolutionary networks: a review.
\newblock {\em J. R. Soc. Interface}, 5(20):259--271, 2008.

\bibitem{funk2010modelling}
Sebastian Funk, Marcel Salath{\'e}, and Vincent~AA Jansen.
\newblock Modelling the influence of human behaviour on the spread of
  infectious diseases: a review.
\newblock {\em J. R. Soc. Interface}, 7(50):1247--1256, 2010.

\bibitem{pastor2015epidemic}
Romualdo Pastor-Satorras, Claudio Castellano, Piet Van~Mieghem, and Alessandro
  Vespignani.
\newblock Epidemic processes in complex networks.
\newblock {\em Rev. Mod. Phys.}, 87(3):925, 2015.

\bibitem{gross2006epidemic}
Thilo Gross, Carlos J~Dommar D'Lima, and Bernd Blasius.
\newblock Epidemic dynamics on an adaptive network.
\newblock {\em Phys. Rev. Lett.}, 96(20):208701, 2006.

\bibitem{shaw2008fluctuating}
Leah~B Shaw and Ira~B Schwartz.
\newblock Fluctuating epidemics on adaptive networks.
\newblock {\em Phys. Rev. E}, 77(6):066101, 2008.

\bibitem{lagorio2011quarantine}
C~Lagorio, Mark Dickison, F~Vazquez, Lidia~A Braunstein, Pablo~A Macri,
  MV~Migueles, Shlomo Havlin, and H~Eugene Stanley.
\newblock Quarantine-generated phase transition in epidemic spreading.
\newblock {\em Phys. Rev. E}, 83(2):026102, 2011.

\bibitem{shaw2010enhanced}
Leah~B Shaw and Ira~B Schwartz.
\newblock Enhanced vaccine control of epidemics in adaptive networks.
\newblock {\em Phys. Rev. E}, 81(4):046120, 2010.

\bibitem{zanette2008infection}
Dami{\'a}n~H Zanette and Sebasti{\'a}n Risau-Gusm{\'a}n.
\newblock Infection spreading in a population with evolving contacts.
\newblock {\em J. Biol. Phys.}, 34(1-2):135--148, 2008.

\bibitem{risau2009contact}
Sebasti{\'a}n Risau-Gusm{\'a}n and Dami{\'a}n~H Zanette.
\newblock Contact switching as a control strategy for epidemic outbreaks.
\newblock {\em J. Theor. Biol.}, 257(1):52--60, 2009.

\bibitem{fefferman2007disease}
NH~Fefferman and KL~Ng.
\newblock How disease models in static networks can fail to approximate disease
  in dynamic networks.
\newblock {\em Phys. Rev. E}, 76(3):031919, 2007.

\bibitem{schwarzkopf2010epidemic}
Yonathan Schwarzkopf, Attila R{\'a}kos, and David Mukamel.
\newblock Epidemic spreading in evolving networks.
\newblock {\em Phys. Rev. E}, 82(3):036112, 2010.

\bibitem{van2010adaptive}
Sven Van~Segbroeck, Francisco~C Santos, and Jorge~M Pacheco.
\newblock Adaptive contact networks change effective disease infectiousness and
  dynamics.
\newblock {\em PLoS Comput. Biol.}, 6(8):e1000895, 2010.

\bibitem{valdez2012intermittent}
LD~Valdez, Pablo~A Macri, and Lidia~A Braunstein.
\newblock Intermittent social distancing strategy for epidemic control.
\newblock {\em Phys. Rev. E}, 85(3):036108, 2012.

\bibitem{guo2013epidemic}
Dongchao Guo, Stojan Trajanovski, Ruud van~de Bovenkamp, Huijuan Wang, and Piet
  Van~Mieghem.
\newblock Epidemic threshold and topological structure of
  susceptible-infectious-susceptible epidemics in adaptive networks.
\newblock {\em Phys. Rev. E}, 88(4):042802, 2013.

\bibitem{lydie2004mobility}
Nathalie Lydi{\'e}, Noah~J Robinson, Benoit Ferry, Evina Akam, Myriam
  De~Loenzien, Severin Abega, Study~Group on~Heterogeneity of HIV Epidemics~in
  African~Cities, et~al.
\newblock Mobility, sexual behavior, and hiv infection in an urban population
  in cameroon.
\newblock {\em JAIDS-J Acq Imm Def}, 35(1):67--74, 2004.

\bibitem{mari2011modelling}
Lorenzo Mari, Enrico Bertuzzo, Lorenzo Righetto, Renato Casagrandi, Marino
  Gatto, Ignacio Rodriguez-Iturbe, and Andrea Rinaldo.
\newblock Modelling cholera epidemics: the role of waterways, human mobility
  and sanitation.
\newblock {\em J. R. Soc. Interface}, page rsif20110304, 2011.

\bibitem{wu2010evolution}
Bin Wu, Da~Zhou, Feng Fu, Qingjun Luo, Long Wang, and Arne Traulsen.
\newblock Evolution of cooperation on stochastic dynamical networks.
\newblock {\em PLoS ONE}, 5:e11187, 2010.

\bibitem{wu2011evolutionary}
Bin Wu, Da~Zhou, and Long Wang.
\newblock Evolutionary dynamics on stochastic evolving networks for
  multiple-strategy games.
\newblock {\em Phys. Rev. E}, 84(4):046111, 2011.

\bibitem{Wu20160282}
Bin Wu, Jordi Arranz, Jinming Du, Da~Zhou, and Arne Traulsen.
\newblock Evolving synergetic interactions.
\newblock {\em J. R. Soc. Interface}, 13(120), 2016.

\bibitem{gardiner1985handbook}
Crispin~W Gardiner.
\newblock {\em Handbook of stochastic methods}, volume~4.
\newblock Springer-Verlag, Berlin, 1985.

\bibitem{pastor2001epidemic}
Romualdo Pastor-Satorras and Alessandro Vespignani.
\newblock Epidemic spreading in scale-free networks.
\newblock {\em Phys. Rev. Lett.}, 86(14):3200, 2001.

\bibitem{albert2002statistical}
R{\'e}ka Albert and Albert-L{\'a}szl{\'o} Barab{\'a}si.
\newblock Statistical mechanics of complex networks.
\newblock {\em Rev. Mod. Phys.}, 74(1):47, 2002.

\bibitem{eubank2004modelling}
Stephen Eubank, Hasan Guclu, VS~Anil Kumar, Madhav~V Marathe, Aravind
  Srinivasan, Zoltan Toroczkai, and Nan Wang.
\newblock Modelling disease outbreaks in realistic urban social networks.
\newblock {\em Nature}, 429(6988):180--184, 2004.

\bibitem{guerra2010annealed}
Beniamino Guerra and Jes{\'u}s G{\'o}mez-Garde{\~n}es.
\newblock Annealed and mean-field formulations of disease dynamics on static
  and adaptive networks.
\newblock {\em Phys. Rev. E}, 82(3):035101, 2010.

\bibitem{pacheco2006coevolution}
Jorge~M Pacheco, Arne Traulsen, and Martin~A Nowak.
\newblock Coevolution of strategy and structure in complex networks with
  dynamical linking.
\newblock {\em Phys. Rev. Lett.}, 97(25):258103, 2006.

\bibitem{durrett2010probability}
Richard Durrett.
\newblock {\em Probability: Theory and Examples}.
\newblock Duxbury Press, Belmont, CA, USA, 2005.

\bibitem{taylor:TPB:2006}
C.~Taylor and M.~A. Nowak.
\newblock Evolutionary game dynamics with non-uniform interaction rates.
\newblock {\em Theor. Popul. Biol.}, 69:243--252, 2006.

\bibitem{gao1992disease}
Linda~Q Gao and Herbert~W Hethcote.
\newblock Disease transmission models with density-dependent demographics.
\newblock {\em J. Math. Biol.}, 30(7):717--731, 1992.

\bibitem{vazquez2015rescue}
F~Vazquez, MA~Serrano, and M~San Miguel.
\newblock Rescue of endemic states in interconnected networks with adaptive
  coupling.
\newblock {\em arXiv preprint arXiv:1511.05606}, 2015.

\bibitem{graser2011separatrices}
Oliver Gr{\"a}ser, PM~Hui, and C~Xu.
\newblock Separatrices between healthy and endemic states in an adaptive
  epidemic model.
\newblock {\em Physica A}, 390(5):906--913, 2011.

\bibitem{tunc2014effects}
Ilker Tunc and Leah~B Shaw.
\newblock Effects of community structure on epidemic spread in an adaptive
  network.
\newblock {\em Phys. Rev. E}, 90(2):022801, 2014.

\bibitem{liggett2013stochastic}
Thomas~M Liggett.
\newblock {\em Stochastic interacting systems: contact, voter and exclusion
  processes}, volume 324.
\newblock Springer-Verlag, New York, 1999.

\bibitem{naasell1999quasi}
Ingemar N{\aa}sell.
\newblock On the quasi-stationary distribution of the stochastic logistic
  epidemic.
\newblock {\em Math. Biosci.}, 156(1):21--40, 1999.

\bibitem{zhou2010evolutionary}
Da~Zhou, Bin Wu, and Hao Ge.
\newblock Evolutionary stability and quasi-stationary strategy in stochastic
  evolutionary game dynamics.
\newblock {\em J Theor. Biol.}, 264(3):874--881, 2010.

\bibitem{ohtsuki2006simple}
Hisashi Ohtsuki, Christoph Hauert, Erez Lieberman, and Martin~A Nowak.
\newblock A simple rule for the evolution of cooperation on graphs and social
  networks.
\newblock {\em Nature}, 441(7092):502--505, 2006.

\end{thebibliography}

\newpage
\section*{Figures}

\begin{figure}[htbp]
   \centering
   \includegraphics[width=6in]{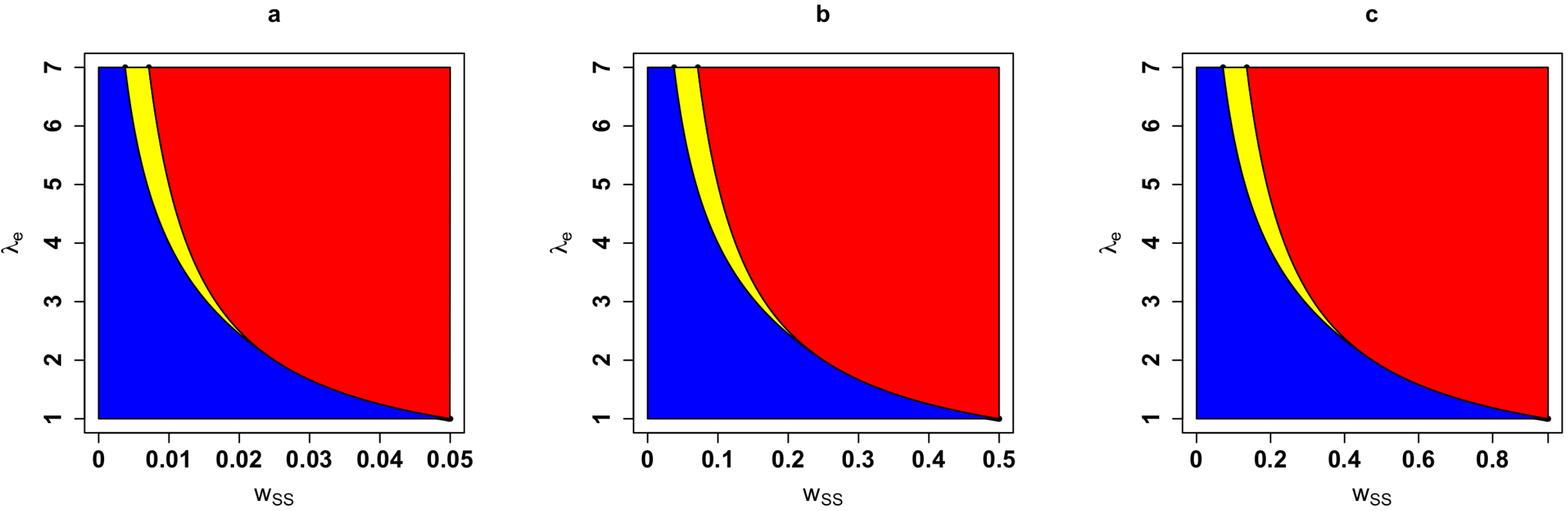}
\caption{SS control, \emph{i.e.}, decreasing the breaking probability between susceptibles.
We start from the uniform case $w_{SS}=w_{SI}=w_{II}$. Here the disease is solely controlled
by increasing the duration time of the social ties between susceptibles, \emph{i.e.}
$w_{SS}<w_{II}=w_{SI}$.
These phase diagrams are similar for all the reference uniform interaction rates:
\emph{i)} for small $\lambda_e$, the SS control makes the disease change from endemic state (red) directly to extinction state (blue);
\emph{ii)} for large $\lambda_e$, the SS control can still eradicate the disease,
but the phase diagram has to pass from endemic to bistable state (yellow) and finally to extinction (blue).
}
   \label{fig:wsi=wii}
\end{figure}

\begin{figure}[htbp]
   \centering
   \includegraphics[width=6in]{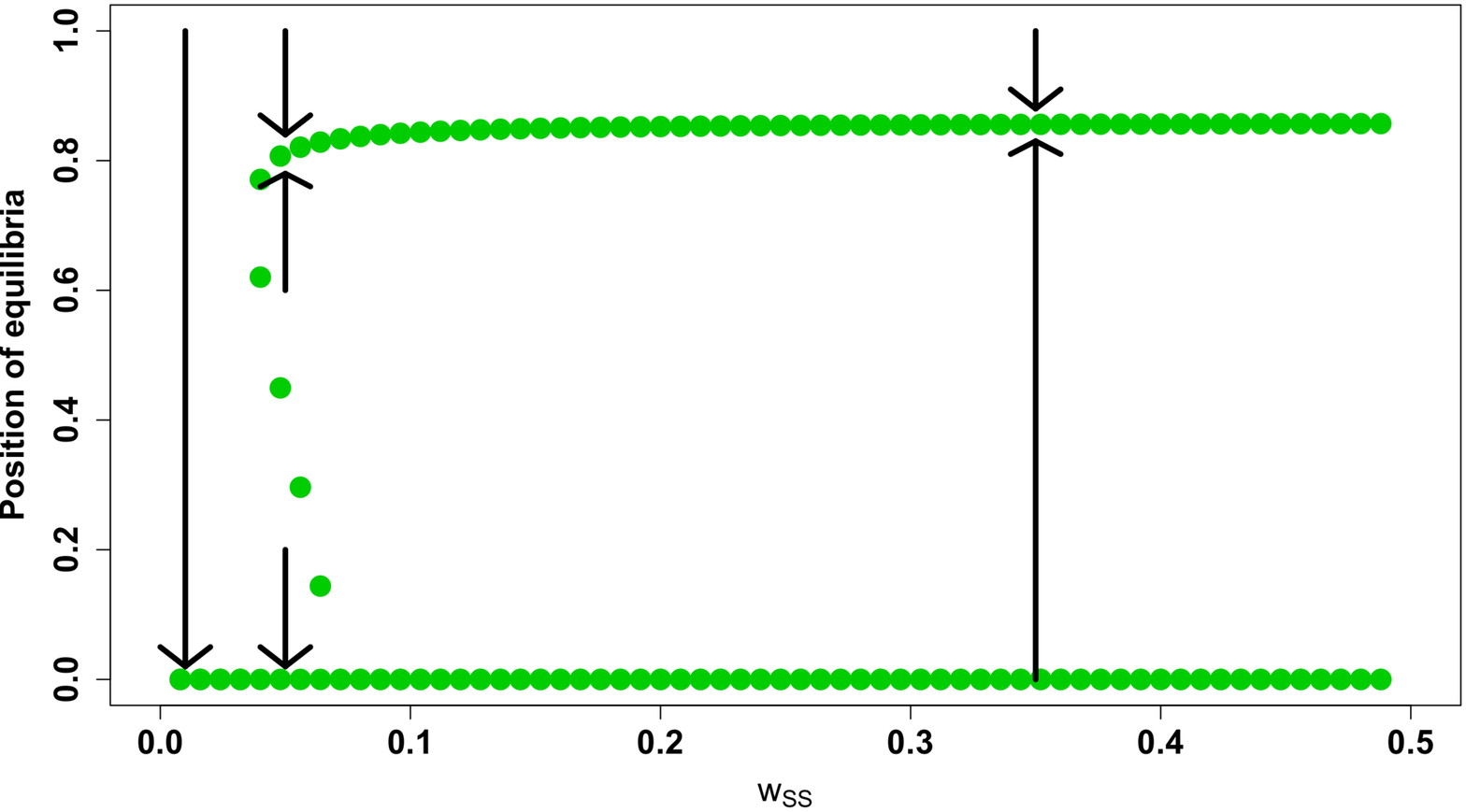}
\caption{Equilibrium position of infected fraction as a function of $w_{SS}$.
Here $w_{SI}=w_{II}=0.5$, and $\lambda_e=7$.
Increasing the interaction time between susceptibles (\emph{i.e.} decreasing $w_{SS}$)
effectively eradicates the disease.
In particular, for $0.033<w_{SS}<0.08$ (bistability),
the disease dies out provided the initial infection is few in number.
Even when the initial number of infection is large,
the final level of infection is still lower than the case with $0.08<w_{SS}<0.5$.
For $w_{SS}<0.033$, the disease is eradicated no matter what the initial state is.
}
   \label{fig:4}
\end{figure}

\begin{figure}[htbp]
\centering
\includegraphics[width=6in]{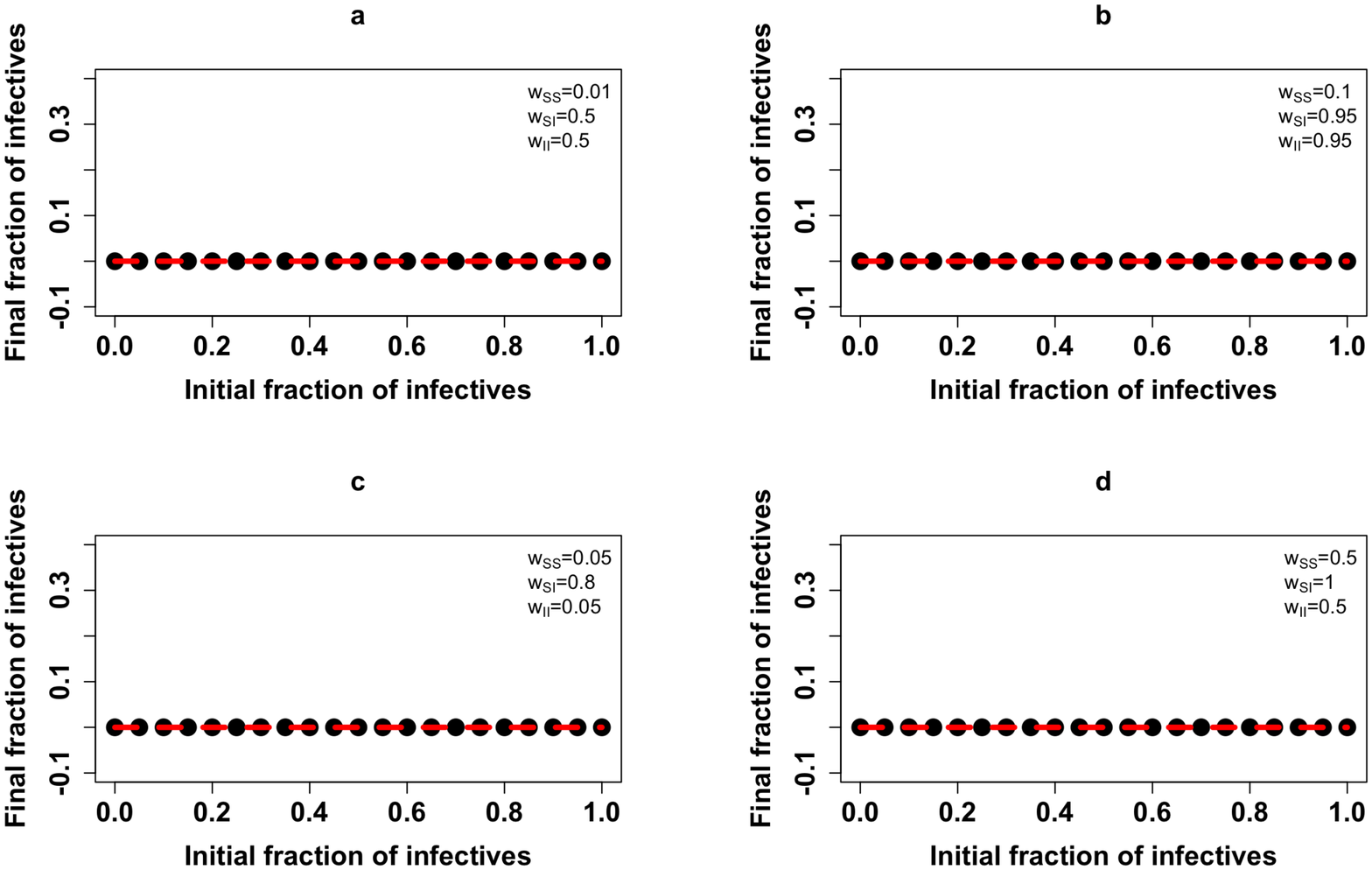}
\caption{Extinction.
The discrete points are obtained from simulations, and the red lines represent the analytical results based on time scale separation.
All of the four panels show that the disease dies out no matter how many infected individuals are present in the beginning.
The parameters in all the four panels are from the blue regions in Figs \ref{fig:wss=wii} and \ref{fig:wsi=wii}.
Thus they are consistent with the analytical predictions.
(Common parameters:  $\lambda_e=1.5$, $N=100$, $\alpha=0.01$.)
}
\label{fig:si:1}
\end{figure}

\begin{figure}[htbp]
\centering
\includegraphics[width=6in]{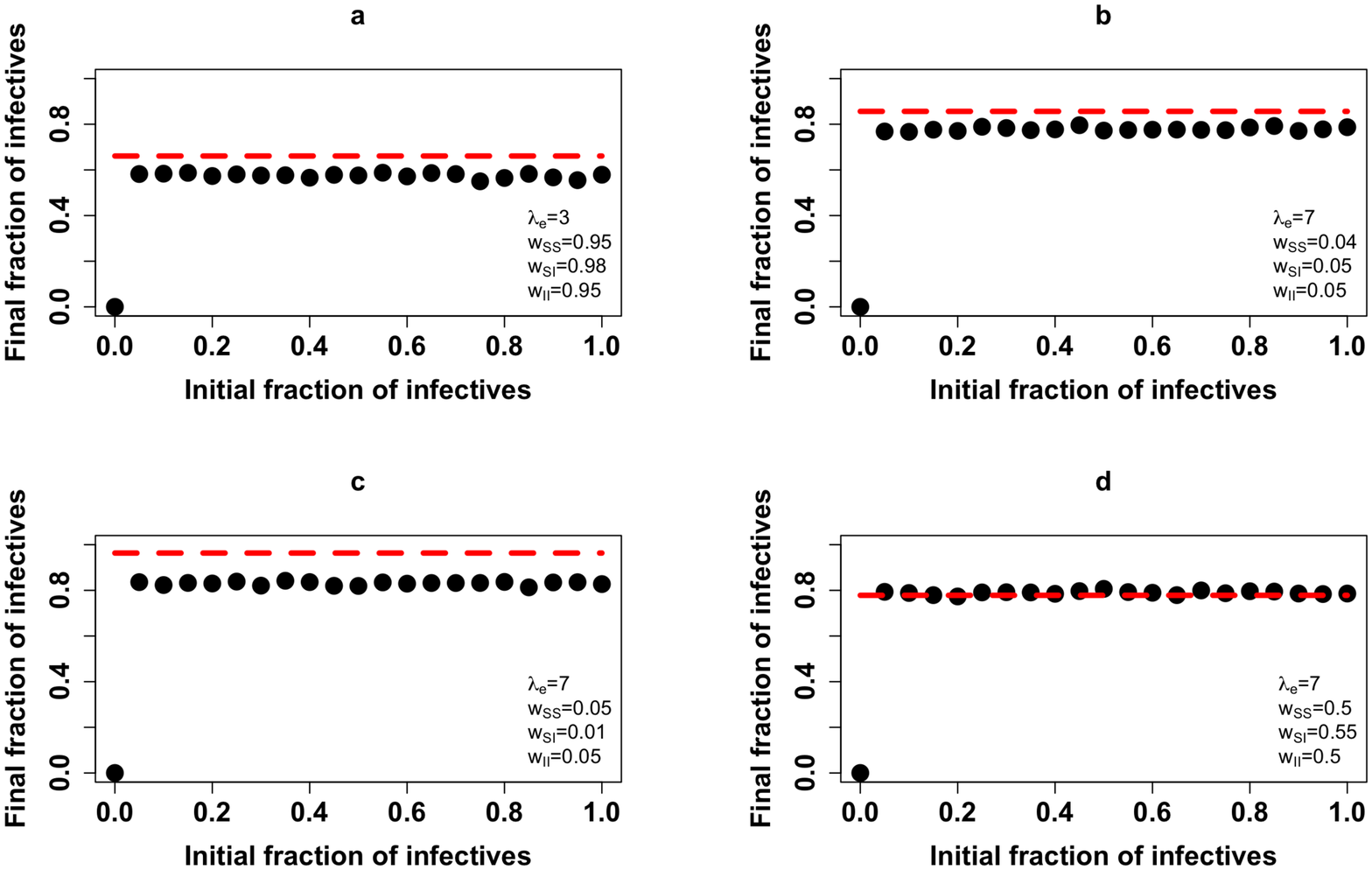}
\caption{Endemic.
The simulation results (discrete points) show that
no matter what the initial fraction of infective individuals is
(except for the zero case which is an absorbing state of the agent-based model),
the population would end up with a constant positive fraction of infected individuals.
However, the analytical predictions would somewhat over-estimate the simulation results.
This is mainly because the running time for the simulation is beyond the quasi-stationary time scale
of our theoretical analysis.
Parameters are from the red regions in Figs \ref{fig:wss=wii} and \ref{fig:wsi=wii}.
(Common parameters: $N=100$, $\alpha=0.01$.)
}
\label{fig:si:3}
\end{figure}

\begin{figure}[htbp]
\centering
\includegraphics[width=6in]{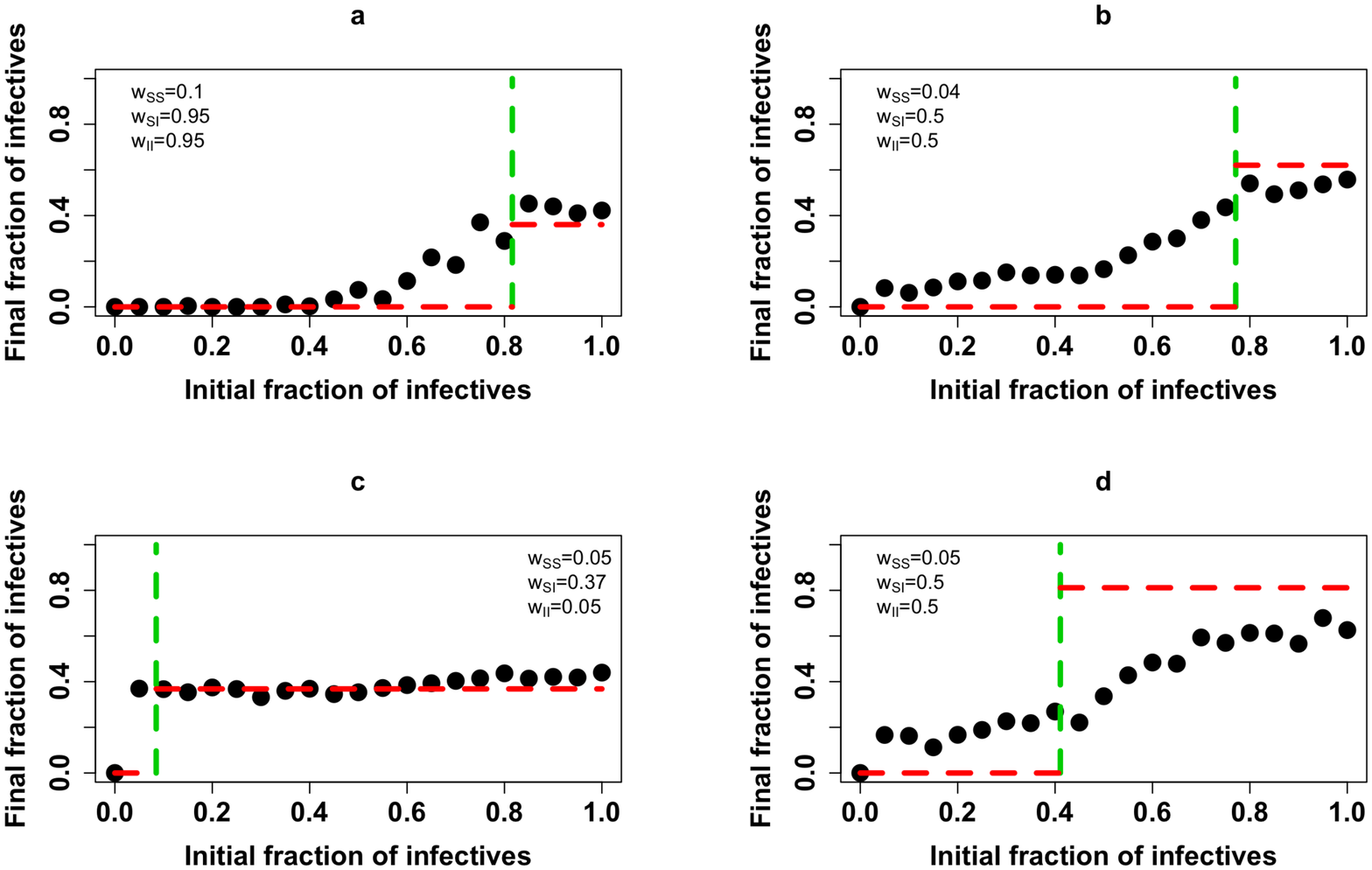}
\caption{Bistability.
Discrete points represent simulation results.
Dashed lines are via the analytical approximations:
Red horizontal lines represent the stable fixed points,
whereas the blue vertical lines represent the unstable fixed point (critical value).
By comparison, the simulation results show qualitative agreement with the analytical predictions.
In particular,
the critical initial fraction of infected individuals, ensuring a dramatic outbreak of epidemics, is consistent with the
unstable fixed point predicted by the analytical result.
However, disagreements are present,
where analytical results underestimate the final fraction of infective individuals
when the infective individuals are rare initially.
Noteworthy, despite of this quantitative inconsistency,
the salient feature of the bistable dynamics is still captured by the analytical predictions.
Parameters are from the yellow region in Figs \ref{fig:wss=wii} and \ref{fig:wsi=wii}.
(Common parameters:  $\lambda_e=7$, $N=100$, $\alpha=0.01$.)
}
\label{fig:si:2}
\end{figure}

\begin{figure}[htbp]
\centering
\includegraphics[width=6in]{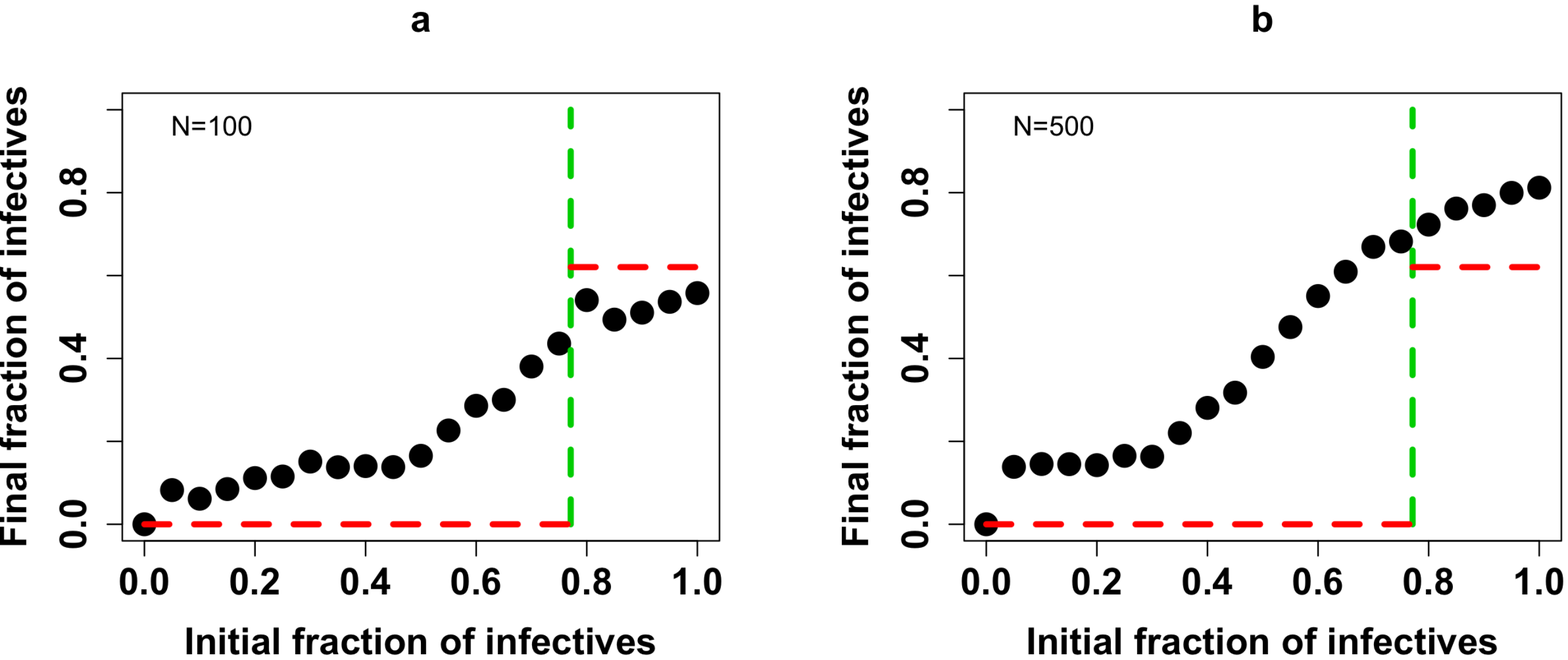}
\caption{The effect of the population size on the accuracy of the analytical approximation.
Population size with $N=100$ is enough to capture the salient feature of the bistability,
compared with $N=500$.
(Common parameters:  $\lambda_e=7$, $w_{SS}=0.04$, $w_{SI}=w_{II}=0.5$ and $\alpha=0.01$.)
}
\label{fig:si:4}
\end{figure}


\end{document}